\documentstyle[prd,aps]{revtex}
\begin{document}
\textheight=640pt
\textwidth=440pt
\topmargin=-25pt
\leftmargin=-25pt
\hoffset -0.4cm

\title{A Model of Evolution with Interaction Strength}
\author{W. Li\thanks{E-mail: liw@iopp.ccnu.edu.cn} 
    and X. Cai\thanks{E-mail: xcai@wuhan.cngb.com} \\ 
   \footnotesize \sl Institute of Particle Physics, Hua-zhong Normal
                     University, Wuhan, 430079, P. R. China}

\maketitle
\vskip 0.5cm
\begin{abstract}
Interaction strength, denoted by $\alpha_{I}$, is introduced in a
model of evolution in $d$-dimension space. It is realized by imposing a constraint 
concerning $2d$ differences of fitnesses between that of any extremal site and those
of its $2d$ nearest neighbours at each time step in the evolution of the
model.  For any given $\alpha_{I}(0< \alpha_I \leq 1)$  
the model can self-organize to a critical state. Two exact equations found in
Bak-Sneppen model still hold in our model for different $\alpha_{I}$. Simulations of
one- and two-dimensional models for ten different values of $\alpha_{I}$ are
given. It is found that self-organized threshold, $f_{c}$, decreases with
$\alpha_{I}$ increasing. It is also shown that the critical exponent, $\gamma$, 
and two
basic exponents, $\tau$, avalanche distribution, and $D$, avalanche dimension,
are $\alpha_{I}$ dependent.
\end{abstract} 
\vskip 0.2cm

\mbox{} \hspace{1.225cm}PACS number(s): 87.10.+e, 05.40.+j, 64.60.Lx

\twocolumn
\hoffset 0.5cm
\rightskip -0.2cm
The concept of self-organized criticality (SOC) concerns the spatiotemporal 
complexity in the systems that contain information over a wide range of
length and time scales [1,2]. It implies that through a dynamical process
a system can start in a state with uncorrelated behavior and end up in a
complex critical state with a high degree of correlation. This concept and
the prototype model, sandpile model, are proposed by Bak, Tang and
Wiesenfeld [3] in 1987. Self-organized criticality is so far the only known
general mechanism to generate complexity [4], and hence the one trying to 
understand why nature is complex, not simple. 

Systems which can exhibit SOC are common in physics, geography, biology, and
even social sciences such as economy. Such kinds of complex phenomena are ubiquitous 
in macroscopic world. Recently, it has been proposed by Meng $et$ $al.$ [5] that
SOC may exist in microscopic systems---at the level of quarks and gluons, as
well as in macroscopic world. 

Evidence from the biology has suggested [6,7] that the ecology of interacting
species had self-organized to a critical state. In 1990, Bak, Chen, and
Crutz [8] created a cellular automaton simulating a society of living
organisms operates at, or very close to, the critical state when driven by
random mutations. However, the model is very sensitive in the sense that
small modifications of the details may drive it away from the critical
state. NKC model proposed by Kauffman and Johnsen [9] can exhibit a
transition from order to disorder, but the criticality emerged in the
system is obtained through parameter tuning, not self-organizing. 
In 1993 [10], a simple model of evolution, Bak-Sneppen model, was introduced by
Bak and Sneppen. Instead of considering the evolution on the individual 
level they present the coevolution of species on a coarsed-grained scale.
In their model the whole species is represented by a single fitness, i.e., a
random number chosen arbitrarily from a uniform distribution between zero and $1$.
And mutations correspond to updating the fitnesses of a given extremal site
and its two nearest neighbours with three new random numbers chosen from the same flat
distribution between zero and $1$. Such model of an evolving biology can
self-organize to a critical steady state during which all sizes of
avalanches can occur. Most important of all, their model can exhibit
punctuated equilibrium behavior observed in biology. Two exact equations
describing self-organization and the average avalanche size behavior
respectively were found later [11,12]. A hierarchical structure of avalanches 
were also observed in B-S model. It was even proposed by Ref. [2] that the
formation of fractal structures, the appearance of $\frac {1} {f}$ noise,
diffusion with anomalous Hurst exponents, Levy flights, and punctuated
equilibria can all be related to the same underlying avalanche dynamics.

Our model of evolution is based on B-S model, but differs decisively in
driving mechanism of interaction between neighboring species. We also consider
coevolution of an interacting species system and each species is represented
by a single fitness, i.e., a random number chosen arbitrarily from a flat
distribution between zero and $1$. But when considering the interaction 
between neighboring species we impose a constraint concerning the
differences of fitnesses between that of the extremal species and those
of its nearest neighbours at each time step. Before knowing how the constraint is imposed let us 
take a first look at another case of evolution, a non-interactive biology.        
In a non-interactive biology each species would tend to evolve towards a
stable state where the fitness of each species approaches $1$, but the
evolution process is extremely slow. If we use $\alpha_{I}$ to denote
$\underline {interaction \\\ strength}$, which represents 
the degree of interaction between the extremal site and its nearest
neighboring species, it is natural to let $\alpha_{I}$ be $1$ in B-S model
and $\alpha_{I}$ be zero in the non-interactive biology. If so, it seems that
these two cases of $\alpha_{I}$ correspond to two extremal cases of interaction
. Then, if $\alpha_{I}$ is allowed to take any value between zero and $1$, 
what can we do with our model? Thus, several questions arise there:

1) What does interaction strength mean for an evolution model?

2) How to present the definition of interaction strength $\alpha_{I}$ and 
impose it into a model of evolution?

3) If the two questions are solved, then are the model of evolution and
some features of it affected by the values of interaction strength?

It will be shown that these questions could be solved successfully in the
following text. 

Our model is intended to consider the evolution of an ecosystem which
consists of a number of species. Followed the ideas proposed in Ref. [10] each
species is represented by a single fitness. The fitness may represent
population of a whole species or living capability of the species.
A high fitness of a certain species may imply that number of the species 
is immense or its living capability is very great, and vice versa. 
Change of fitness of a species may imply change of number of the species
or change of its living capability.  So it is natural to expect that a species 
with a low fitness is more likely to change, namely mutation in biology,  
in order to live better and/or longer in nature. Only through mutations the 
species with a bad fitness can have the chances of choosing a better fitness 
in order to avoid extinction. Furthermore, the fitness of each species 
is affected by other species which are also parts of the ecosystem. 
Any adaptive change of any species may change the fitness and the fitness 
landscape of its coevolutionary parts coupled in the same ecosystem. 
So the species may interact with each other through, say, a food chain. Hence,
the species with a high fitness may live well and comfortably unless its 
bad neighbours are going to mutate. 

Our model is defined and simulated through the following items:

(1) A number of, say, $L^{d}$ species are located on a $d$-dimensional lattice
of linear size $L$. Initially, random numbers chosen arbitrarily from a 
uniform distribution between zero and $1$, $p(f)$, are assigned independently 
to each species. At each time step,

(2) choose the extremal site, that is, the species with the lowest fitness,
$f_{\rm min}$, among all the species and update it by assigning a new random number also 
chosen from $p(f)$ to it and

(3) mutate those of its $2d$ neighbours whose fitnesses satisfy the
constraint $f_i-f_{\rm min} < \alpha$ by assigning new random numbers between zero and $1$
to them ($f_{\rm i}$ denotes the fitness of the $i$th
nearest neighbours.), $\alpha$ is a parameter between zero and $1$ and is fixed 
for a given model. 

It should be pointed out that different values of $\alpha$ correspond to 
different versions of the model, even to say, different models. Consider
two special values of $\alpha$, 0 and $1$. For $\alpha=0$ the model returns to
the non-interactive biology since difference of fitness between those of
any two neighbours is always greater than zero, so in the model in which
$\alpha=0$ none of
the neighbours of any extremal species will be chosen for updating at each
time step. This 
is the case where there is no interaction between neighbours. For $\alpha=1$
the model returns to a $d$-dimensional B-S model. It is because that difference of 
fitness between those of any two neighbours is 
less than $1$ so at each time step all the $2d$ nearest neighbours of
the extremal site will be chosen for updating.

For a given $\alpha$ let the model evolve from the beginning when the first
extremal species is chosen and we determine how many of its $2d$ neighbours 
will be chosen, according to the constraint on difference of fitness, for
updating at the same time step. The updating process, i.e., the evoultion of
the model, continues indefinitely. For a given $\alpha$ satisfying $0 < \alpha <
1$ the updating of the $2d$ nearest neighbours of the extremal site at each
time step will have many kinds of probabilities. For instance, maybe
none of the $2d$ neighbours will be updated, maybe half of the $2d$ neighbours
will be chosen for updating, $etc$. If we do not distinguish the neighbours
when we only care the number of the updated neighbours at each time step
it is straightforward that the updating of the neighbours will have $2d+1$ 
probabilities. Say, at some time step, an extremal site and $m$ of its $2d$
nearest neighbours are chosen for updating according to the constraint. Such
an updating is called by us a $m$-event. As the evolution goes on we can
observe many kinds of such events, during which $m$ can be different. If the
evolution time is large enough we can obtain $2d+1$ kinds of events during
which $m$ spans through $0$ and $2d$. If we define $P_{d}(m)$ the probability of
$m$-event among all events during time period $T$, i.e.,

\begin{equation}
P_{d}(m)=\frac {N(m)} {N_T} ,
\end{equation}                
  
\noindent
 where $N(m)$ is the number of $m$-event during the time period $T$ and
$N_T$, the total number of all events, and $d$ denotes dimension. In the $T$ limit ($T \gg L^{d}$)
we will obtain the distribution of $m$-event, that is,
\begin{equation}
P_{d}(m)=\lim_{T \rightarrow \infty} \frac {N(m)} {N_T} .
\end{equation}
\noindent
And $P_{d}(m)$ should satisfy the normalization,
\begin{equation}
\sum\limits_{m=0}^{m=2d}P_d(m)=1 .
\end{equation} 

Next we will present the definition of interaction strength $\alpha_{I}$ 
related to $P_{d}(m)$,

\begin{equation}
\alpha_{I}=\lim_{T \rightarrow \infty} \frac {1} {2d}
\sum\limits_{m=0}^{2d} m \, P_{d}(m) .
\end{equation}

\noindent
One can see that $\alpha_{I}$ is actually the statistical ratio of 
number of updated neighbours among $2d$ ones during the evolution.
This can be easily seen from two extremal cases when $\alpha=0$
and $\alpha=1$. For $\alpha=0$, $\alpha_{I}=0$, this is because $P_{d}(0)=1
$ and $P_{d}(m)=0$ for $0 < m \leq 2d$; for $\alpha=1 $, $\alpha_{I}=1$,
this is because $P_{d}(2d)=1$ and $P_{d}(m)=0$ for $0 \leq m < 2d$. That
is, in the nointeractive biology interaction strength $\alpha_{I}$
is zero while in the B-S model $\alpha_{I}$ is 1. Hence,
it is natural to expect that the definition of $\alpha_{I}$ can give a good
description of the change of strength of interaction between 
neighboring sites when $\alpha$ changes. So, it is also natural to expect 
that $0< \alpha_{I} <1$
for $0< \alpha <1$, and that $\alpha_{I}$ increases as $\alpha$ increases. There
should also exists one-to-one correspondence between $\alpha_{I}$ and
$\alpha$. Having these in mind we measure the distribution $P_d(m)$ 
and $\alpha_{I}$ for ten
different values of $\alpha$ for one- and two-dimensional models on the
computer. Simulation results are given in Fig. 1 and Fig. 2 respectively. 
Figures (a) and (b) in Fig. 1 show the distribution of $m$-event ($0\leq m \leq
2d$) for ten different 
values of $\alpha$ in one- and two-dimensional models respectively.
It is clearly shown in the above figures that $P_{d}(0)$ decreases with 
increase of $\alpha$ and $P_{d}(2d)$, i.e., $P_{d=1}(2)$ for $d=1$ 
and $P_{d=2}(4)$ for $d=2$,
increases as $\alpha$ increases. Note in the figures change of $P_d(m)\,(0<
m< 2d)$ with increase of $\alpha$. Their exists a peak in the curve of 
$\alpha$ dependence of $P_d(m)\,(0< m< 2d)$. This can ensure the
normalization of $P_d(m)$. Two plots in Fig. 2 show the
dependence of $\alpha_{I}$ on $\alpha$ in one- and two-dimensional
models respectively. As shown in Fig. 2 $\alpha_{I}$ almost 
increases linearly as $\alpha$ increases, and most important of all is 
that one $\alpha_{I}$ corresponds to one and only one $\alpha$. 
From here one can clearly see that our
definition of $\alpha_{I}$ can be explicitly related to $\alpha$ and hence
can be put into our model naturally. Relating Fig. 1 and Fig. 2 one can see
that increase of interaction strength increases $P_d(2d)$. That is, 
when $\alpha_I$ increases any extremal site will possibly affect its nearest
neighbours more strongly. Furthermore, different values of $\alpha_I$
correspond to different interaction strength. Generally speaking, larger
$\alpha_I$ represents stronger interaction, and vice versa. Hence, the definition of $\alpha_{I}$ 
can provide a good description for the strength of interaction between 
neighboring sites. It is a good quantity in describing the interaction. In the
following 
text dependence of self-organized threshold and some critical exponents 
on $\alpha_{I}$ will be given. Since $\alpha_{I}$ has one-to-one 
correspondence with $\alpha$ and increases as $\alpha$ increases, 
it is hence convenient  and equivalent
to present the dependence of these quantities on $\alpha$.        

The model is already defined, it is natural to investigate whether 
the model can self-organize to a critical state. That is, we should observe
the ``fingerprint`` of SOC [4]. If so, it is also 
worthwhile to know whether the criticality is sensitive to the 
value of $\alpha$. In addition, the self-organization, referred to a
dynamical process whereby a system starts in a state with uncorrelated
behavior and ends up in a complex state with a high degree of correlation,
of the system and punctuated equilibrium, the most important feature of 
evolution, should also be observed. 

Fig. 3 shows the space-time fractal activity pattern for a one-dimensional
model of size $L=100$ with $\alpha=0.8$. We track the updated sites
at each time step. $S$ and $R$ in the figure denote number of updated time steps 
and location of updated sites respectively. Simulations of one- and
two-dimensional models for different values of $\alpha$ are also done and
exhibit the similar space-time fractal activity. Indeed, spatiotemporal
complexity emerges in our model and its appearance is independent of the 
value of $\alpha$ choosen.

We now present some explanations of the quantities which will appear in the 
following equations for those readers who are not so familar with SOC.
$f_{\rm min}$ denotes the extremal fitness at each time step in the
evolution. 
$G(s)$, the gap appeared in punctuated equilibrium, is an envelope function 
that tracks the increasing peaks in $f_{\rm min}$. Its definition is : 
at time step $s$ the gap $G(s)$ is the maximum of all the minimum 
random numbers chosen, $f_{\rm min}(s^{\prime})$, for all $0 \leq s^{\prime} \leq s$. 
$f_c$ is the value of $G(s)$ at critical state, i.e.,

\begin{equation}
f_c=\lim_{s\rightarrow \infty}G(s).
\end{equation}
\noindent
$S_{G(s)}$ is the size of avalanches correspond to plateus in $G(s)$ during which 
$f_{\rm min}(s) < G(s)$ and 
$\langle S \rangle_{G(s)}$ is the average value of $S_{G(s)}$. 
An avalanche is defined as subsequent mutations below a
certain threshold. Hence with this definition there is a hierarchy of
avalanches each defined by their respective thresholds. So we can have
$f_0$-avalanche where $f_{0}$ is only an auxiliary parameter 
between zero and $1$ to define avalanches. More detailed definition of
$f_0$-avalanche is given in Ref. [5]. The size of an avlanche, $S$,
is the number of subsequent mutations below the threshold. And $\gamma$ 
is a critical exponent which governs the divergence of $S_{G(s)}$ 
when $s$ approaches infinity. And
$n_{\rm cov}$
is the number of sites covered by an avalanche. 
Apparently $n_{\rm cov} \leq (2d+1)S$ ($S$ is the avalanche size) 
in d-dimensional space. $\langle n_{\rm cov} \rangle$ denotes the average 
value of $n_{\rm cov}$. The above defined quantities will have their
counterparts in our model. So we will use the same definitions of these
quantities while make some minor corrections on the symbols of them.  

Following the method used in Ref. [2] we monitor the extremal signal $f_{\rm min}$ 
as a function of $s$ during the transient in the one- and two-dimensional 
models for different values of $\alpha$. Again, we observe Devil's staircase
[4] in all these cases. Fig. 4 shows punctuated equilibrium behavior in
one-dimensional model of size $L=100$ with $\alpha=0.5$. Hence, we can see
punctuated equilibrium does emerge in our model of evolution and 
its emergence is $\alpha$ independent. 

Observations through above simulations suggest that our model 
can self-organize to a critical state. But how to determine 
the self-organized threshold $f_c$ is
still a hard bone to us. Fortunately Ref. [2] provides us a method. 
In Ref. [2] the
author presents two exact equations for Bak-Sneppen model. With a reasonable scaling ansatz
of average size of avalanches they can determine $f_c$ very accurately. 
It is straightforward to expect that the two exact equations found 
in Bak-Sneppen model have the  corresponding ones in our model. 
According  to the derivation of these two equations [2]
we can directly write down two similar exact equations in our model except 
for some minor replacements on the symbols of some quantities. Introduction
of interaction strength $\alpha_{I}$ suggests that some quantities are
$\alpha_{I}$ and hence $\alpha$ dependent, which can be seen from our
simulations: as $\alpha$ increases $f_c$ has a tendcy to decrease. Having
this in mind we make replacements of symbols of some quantities:
\begin{eqnarray}\nonumber
G(s) &\rightarrow G(s,\alpha),\\ \nonumber 
f_{c}   &\rightarrow f_{c}(\alpha), \\ \nonumber
\gamma   &\rightarrow \gamma(\alpha), \\ \nonumber
\tau & \rightarrow \tau(\alpha), \\ \nonumber
D &\rightarrow D(\alpha) .
\end{eqnarray} 

\noindent
Two exact equations are given as below,
\begin{equation}
\frac {{\rm d}G(s,\alpha)} {{\rm d}s}= \frac {1-G(s,\alpha)} {L^{d} \langle S \rangle
_{G(s,\alpha)}}
\end{equation}
\begin{equation}
\frac {{\rm d ln} \langle S \rangle_{f_0}} {{\rm d}f_0}=\frac {\langle n_{\rm cov} \rangle_{f_0}} {1-f_0}
\end{equation}

In order to solve the two equations a scaling ansatz of $\langle S \rangle_{G(s,\alpha)}$
should be given,
\begin{equation}
\langle S \rangle_{G(s,\alpha)} \sim [f_c(\alpha)-G(s,\alpha)]^{-\gamma(\alpha)} .
\end{equation}

Inserting Eq. (8) into Eq. (7) one obtains 
\begin{equation}
\gamma(\alpha)=\lim_{f_0 \rightarrow f_c(\alpha)} \frac {\langle
n_{\rm cov} \rangle_{f_0} 
[f_c(\alpha)-G(s,\alpha)]} {1-f_0} .
\label{gamma}
\end{equation}

Using Eq. (\ref{gamma}) we can determine $f_c(\alpha)$ and $\gamma(\alpha)$ very
accurately. We measure $f_c(\alpha)$ and $\gamma(\alpha)$ in the one- and
two-dimensional models for different values of $\alpha$. It is to our
expectation that
these two quantities are $\alpha$ dependent. Fig. 6 shows the dependence of
$f_c(\alpha)$ on $\alpha$ and figure (a) in Fig. 7 shows that of $\gamma(\alpha)$ on
$\alpha$. In addition, we measure two basic exponents of the model,
$\tau(\alpha)$, avalanche size distribution, and $D(\alpha)$, avalanche
dimension. We find they are also $\alpha$ dependent.  Figures (b) and (c) in
Fig. 7 show their dependence on $\alpha$ respectively.

The following text will present our analysis of the simulations and our
conclusion. Let us first check the ``fingerprint`` of SOC. Fig. 5 shows avalanche distribution $P_{\rm aval}(S)$ for one-dimensional 
model of size
$L=100$ with $\alpha=0.7$. Such distribution can also be found in one- and
two-dimensional models for different $\alpha$. Indeed, power law emerges in
our model and its appearance is $\alpha$ dependent. Then let us come to the
results of $f_c(\alpha)$. Two plots,
(a) and (b), in Fig. 6 show the dependence of $f_{c}(\alpha)$ on $\alpha$ in the one- and
two-dimensional models respectively. It is clearly shown that these two
figures display the similar behaviors of $f_{c}(\alpha)$. 
Firstly, we can see that in both figures $f_{c}(\alpha)$ decreases 
as $\alpha$ increases. It is
not difficult to understand this kind of behavior. As $\alpha$ increases,
i.e., $\alpha_{I}$ increases, the chances for the nearest neighbours of a
given extremal site to be chosen for updating at each time step will be
greater. This can be easily seen from Fig. 1. As shown in Fig. 1 $P_d(2d)$
increases as $\alpha$ increases and reaches $1$ for $\alpha =1$, and
$P_d(0)$ decreases as $\alpha$ increases and reaches zero when $\alpha=1$. 
So, with the
increase of $\alpha$ , that is the increase of $\alpha_{I}$
, more possible neighbours will be involved in the evolution, hence, the
threshold $f_{c}(\alpha)$ will be lowered further. This can be explained in
another way. Compare the values of $f_{c}(\alpha)$ for the same $\alpha$ in
one- and two-dimensional models. Say, compare $f_{c}(1)$ in one
dimensional model with that in two-dimensional model. It has been shown that
$f_{c}(1)$ in the former case is greater than that in the latter case. In one
-dimensional model in which $\alpha =1$ when an extremal site is chosen,
two of its nearest neighbours will also be chosen, for updating at each
time step. While in two-dimensional model with $\alpha=1$, an
extremal site, together with its four nearest neighbours, will be chosen
for updating at each time step. Why we mention this two cases here is just
trying to show that the increase of average number of neighbours involved in the
evolution will lower the self-organized threshold. Thus, one can see, 
the increase of interaction strength will increase the number of
neighbouring sites involved in the evolution and hence lowers down the
value of $f_{c}(\alpha)$. Secondly, it is shown in both figures of Fig.
6 that $f_{c}(\alpha)$ almost decreases linearly as $\alpha$ increases for
$0<\alpha<0.6$ and decreases asymptotically  as $\alpha$ increases for
$\alpha$ between $0.6$ and $1.0$. This implies that effect of the constraint
on the evolution grows implicitly as $\alpha$ increases. When interaction
strength is small the effect is very explcitly. But as interaction strength
increases the effect will be not so explicitly shown. Specifically, we
measure $f_{c}(\alpha)$ when $\alpha =1$ for one- and two-dimensional models.
We find $f_{c}(1)=0.668 \pm 0.001$ for $d=1$ system of size $L=100$ , which is
very close to the value in Refs. [2] and [13], who found $f_{c}(1)=0.660702 \pm 0.00003$;
for $d=2$ system of size $L=20$, $f_{c}(1)=0.334 \pm 0.00006$ which is close
to the corresponding value in Ref. [2], who found $f_{c}(1)=0.328855 \pm
0.000004$. In addition, for $d=1$ we find $f_{c}(0.1)=0.96388 \pm 0.000002$ 
for a system of size $L=100$ and for $d=2$, $f_{c}(0.1)=0.9179 \pm 0.00001$
for a system of size $L=20$.         
            
We also measure $\gamma(\alpha)$ for different $\alpha$ in one- and
two-dimensional models. It is found that $\gamma$ is also $\alpha$
dependent. Dependence of $\gamma(\alpha)$ on $\alpha$ is given in Fig. 7.
In figure (a) in Fig. 7 $\gamma(\alpha)$ first increases and then 
decreases as $\alpha$ increases. For $d=1$, $\gamma(1)=2.6166 \pm 0.00004$,
which is close to the value found in Refs. [2,13,14], who found $\gamma(1)=2.70
\pm 0.01$. For $d=2$, we find $\gamma(1)=2.249 \pm 0.0004$, which is
not in agreement with Ref. [2], who found $\gamma(1)=1.70 \pm 0.01$. Maybe
this is because of our small size of system. If the size is larger the results    
will be more precise.

For different values of $\alpha$ in one- and two-dimensional models we
measure two basic exponents, $\tau(\alpha)$, which characterizes the
distribution of avalanche sizes, and $D(\alpha)$, the avalanche dimension.
The definition of $\tau(\alpha)$ is $P_{\rm aval}(S) \sim
S^{-\tau(\alpha)}$, and that of $D(\alpha)$ : $n_{\rm cov} \sim
S^{D(\alpha)/d}$. Results of these
two basic exponents are given in Fig. 7(b) and Fig. 7(c) respectively. It is shown in
the plots that $\tau(\alpha)$ first decreases as $\alpha$ increases and then 
changes slowly with variation of $\alpha$. Specifically, for $d=1$, we find
$\tau(1)=0.858 \pm 0.001$, which is not in agreement with Ref. [2], who measured
$\tau=1.07 \pm 0.001$, but agrees with Ref. [10], who measured $\tau=0.8 \pm 0.1$.
This is because our size $L=100$ is close to the size $L=64$ in Ref. [10]
but far from the size $L=10^4$ in Ref. [2]. For $d=2$, $\tau(1)=1.131 \pm
0.0005$, which is close to Ref. [2], who measured $\tau=1.245 \pm 0.01$.

Fig. 7(c) shows the dependence of $D(\alpha)$ on $\alpha$ for one- and
two-dimensional models. It can be inferred from the plots
that $D(\alpha)$ first decreases rapidly and then slowly as $\alpha$
increases. For $d=1$, $D(1)=2.4189 \pm 0.0001$, which is close to Ref. [2],
who measured $D=2.43 \pm 0.01$, and for $d=2$, $D(1)=2.94 \pm 0.0001$,
which agrees with Refs. [2,9,10], who measured $D=2.92 \pm 0.02$. It is clearly 
shown that for all $\alpha$ $D(\alpha)$ is larger than the dimension space
$d$.

It should be emphasized that interaction strength, $\alpha_{I}$, is not a
parameter tuning the model to a critical state, since in our model when
a model is chosen the corresponding $\alpha_{I}$ is fixed during the whole 
evolution process. Thus, criticlity emerged in our model is self-organized,
not tuned. And appearance of criticality in our model is independent of
$\alpha_{I}$ chosen, which you can test on your own PC. This can
strongly support the idea that SOC does not depend on the dynamical details 
of the system. That is, self-organized criticality should be universal. Furthermore, self-organized
fractal growth is basically different from growth processes, say, described
by (variants of) the Kardar-Parisi-Zhang (KPZ) equation [15,16]. The KPZ
equation is scale invariant by symmetry, thus the criticality is not
self-organized. SOC cannot, even in principle, be regarded as sweeping a
system through a critical point, which contrasts to the claims in Ref. [17].
SOC should be an attractor of the complex system, but this attractor is 
vastly different from the one found in chaos---"strange attractor".

As shown in this paper and in others [2,4,10], interaction plays a very
important role in the models which exhibit SOC. If there is no interaction
between individuals in a system, the system will evolve towards a frozen
state and the evolution process will be indefinitely long. This is clearly
shown in non-interactive biology in which fitness of each species tends to
be $1$. Our model and simulations of its different versions corresponding to
different degrees of interaction between neighbours imply that only
the coevolutionary system can evolve to a self-organized critical state
despite the fact that interaction strength may be relatively small. Through our
simulations that we have learned it would take longer time for the system to
involve to a critical state when the interaction strength $\alpha_{I}$ is
smaller. It is also worthwhile to perform two simulations of the models
in which $\alpha_{I}$ is very close to zero and $1$ respectively. We can expect
in the former case the model will evolve to a frozen state, while in the
latter case we will approach B-S model.

Another important feature of our model is that fitness itself is directly
involved in the interaction, which is not realized in the B-S model. As a 
coevolutionary system, interaction between any extremal species and
its neighbours should and must involve the features of the 
extremal site and that of the neighbours. Because of model's simplicity
each species has only one feature: fitness, represented by a random number
chosen arbitrarily from a flat distribution between zero and $1$. Hence, this
feature should enter the evolution model. Through a constraint which can
be related to interaction strength the fitness is involved in 
and injected into the evolution process of the system. It is shown in our
model that evolution is indeed an coevolutionary phenomena, which agrees
with Darwin's opinion on evolution of the biology [18].

Thus, in conclusion: 

(1) A simple model of evolution with interaction strength defined and
considered is proposed. The models with different interaction strength
($0< \alpha_I \leq 1$) can self-organize to critical states.

(2) Simulations of one- and two-dimensional model of various degrees of
interaction strength show that $f_{c}(\alpha_{I})$ decreases as
$\alpha_{I}$
increases. It is also shown that $\gamma(\alpha_{I})$, and two basic exponents
, $\tau(\alpha_{I})$ and $D(\alpha_{I})$, are $\alpha_{I}$ dependent.

This work was supported by NSFC in China. The authors thank Prof. T. Meng and other members of our small group on SOC 
for their fruitful discussions during his stay in wuhan.

\vskip 1cm
\begin{center}
\bf Figure Captions:
\end{center}

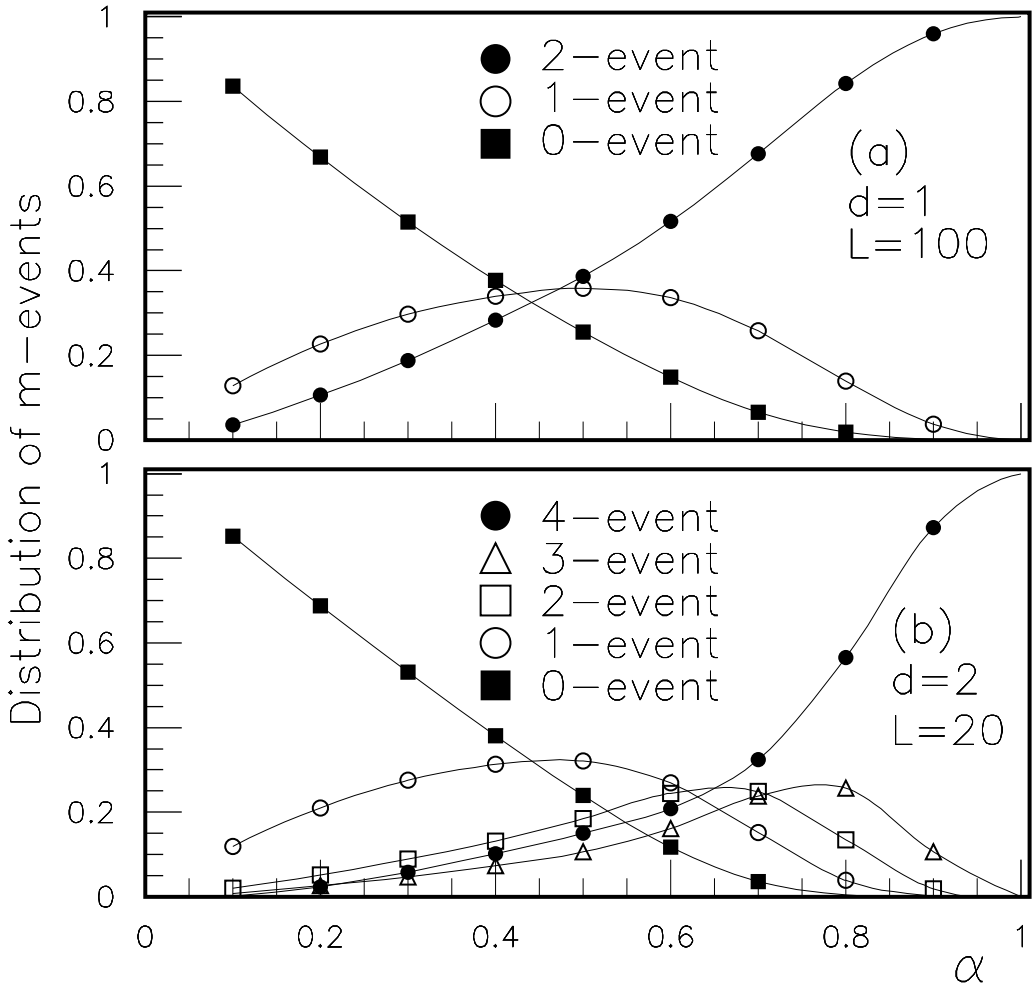
Fig. 1: Distribution of m-event for different $\alpha$ in the
evolution of (a)one-dimensional models of size $L=100$ and (b)two-dimensional
models of size $L=20$.

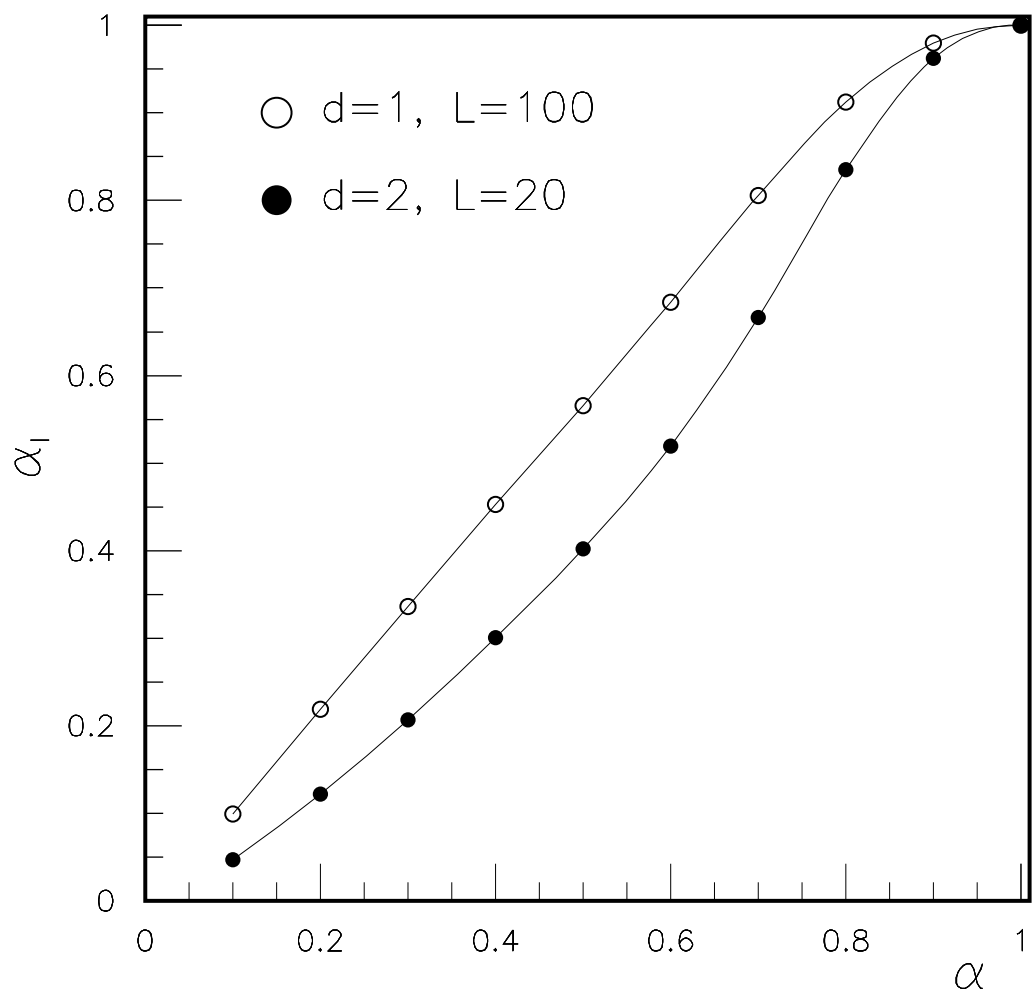
Fig. 2: Dependence of interaction strength $\alpha_{I}$ on $\alpha$
in the evolution of one-dimensional models of size $L=100$ and two-
dimensional models of size $L=20$.

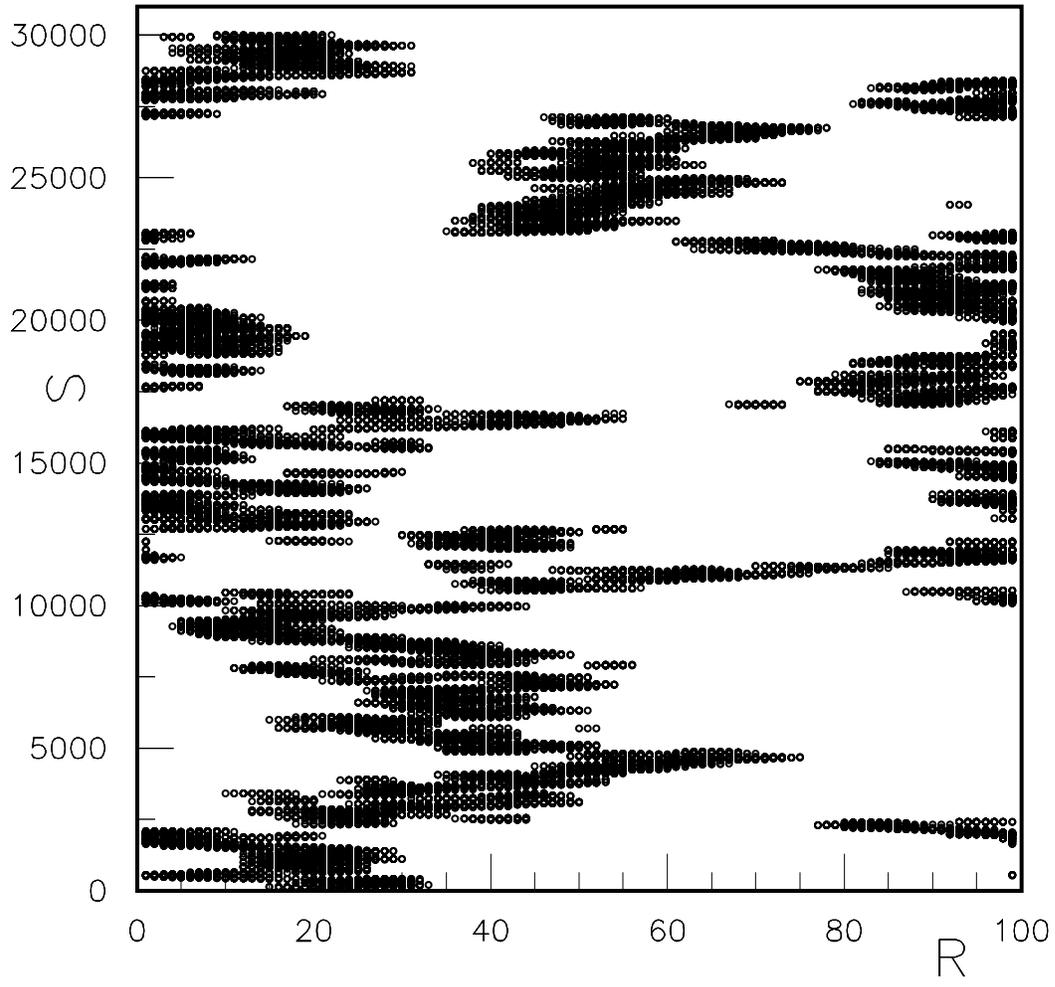
Fig. 3: Space-time fractal activity pattern for one-dimensional evolution 
model
with  $\alpha=0.8$. $S$ is the number of updated steps and $R$ is the
location of updated sites.  

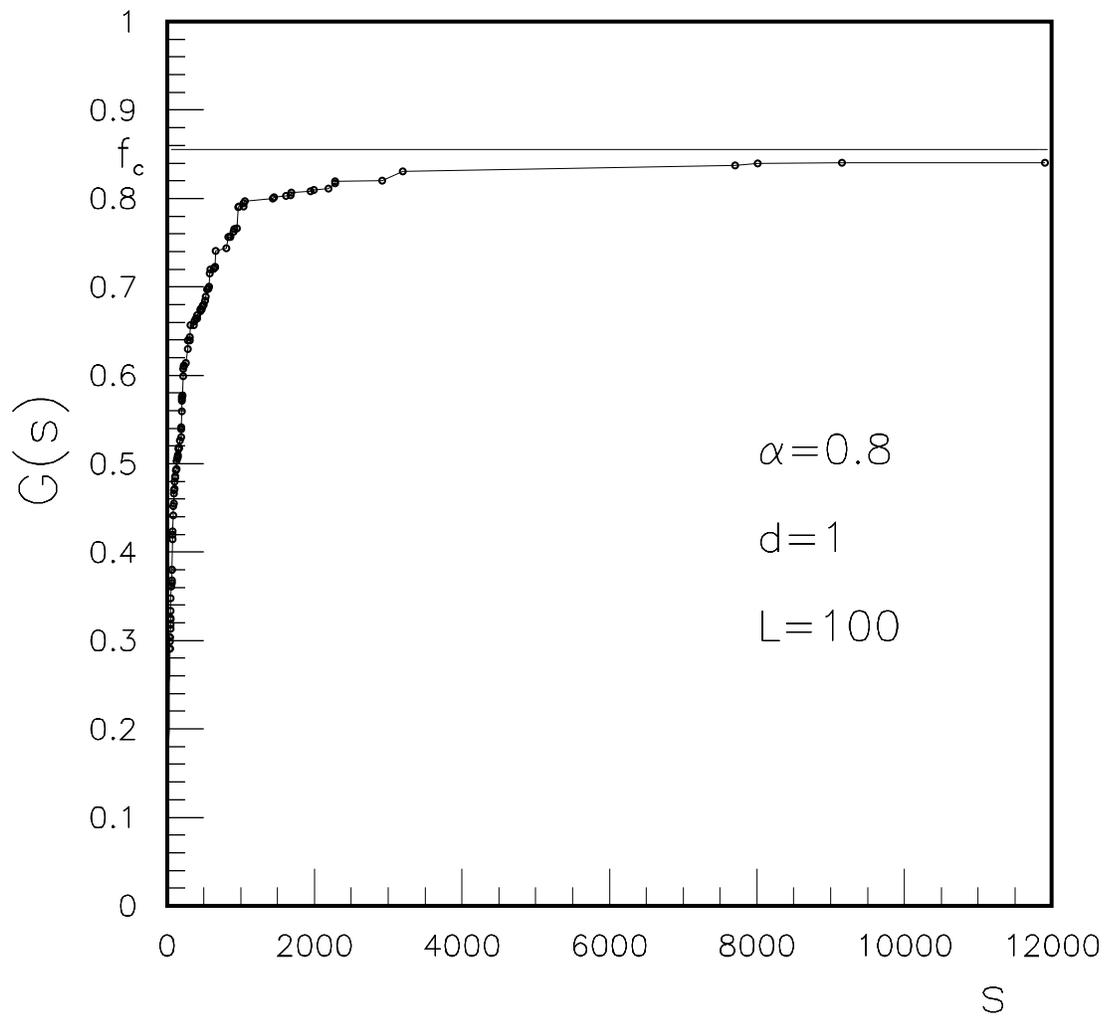
Fig. 4: Punctuated equilibrium behavior emerges in one-dimensional evolution
models of size $L=100$ with $\alpha=0.5$. $G(s)$ is the gap that tracks 
the peaks of extremal signal, $f_{\rm min}$, in the transient. $f_c$ in the plot is about $0.8556$.  

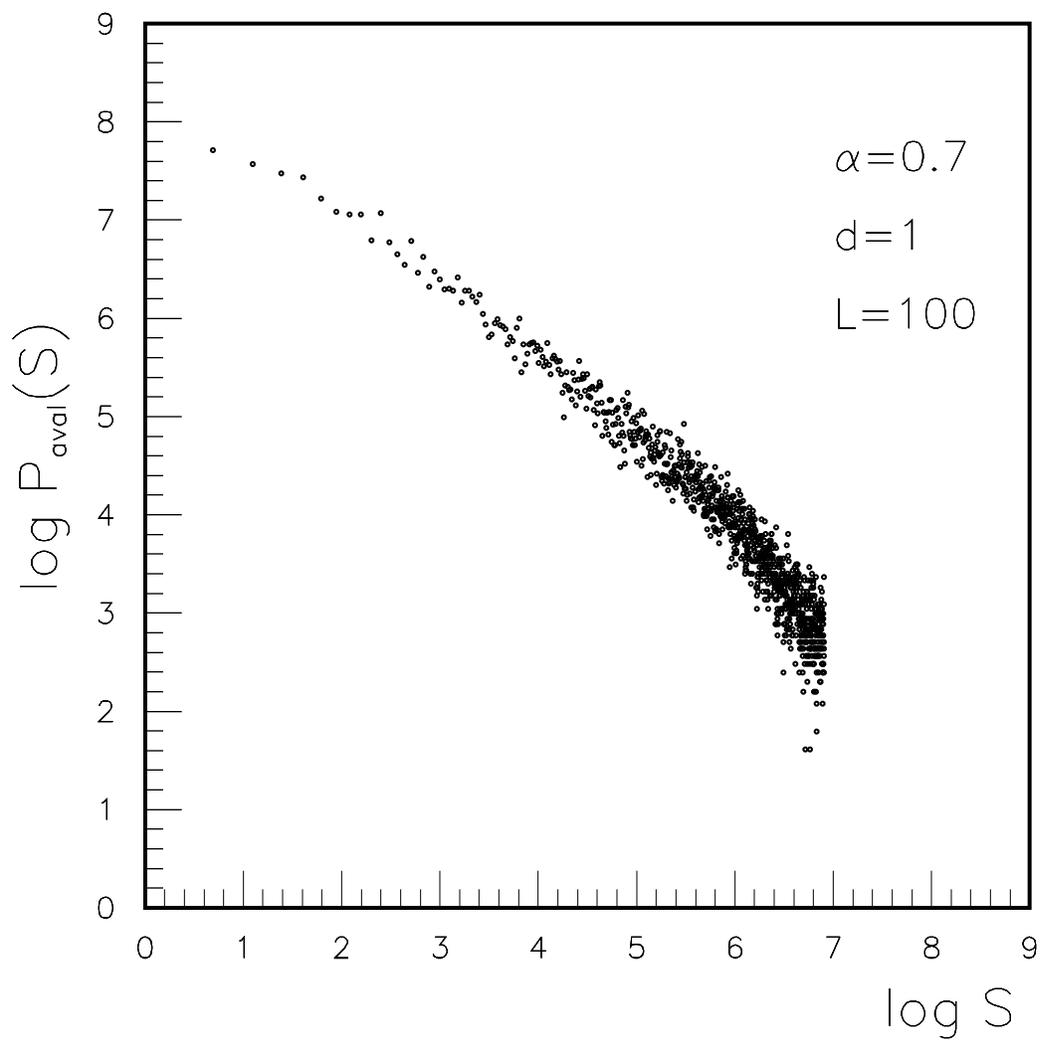
Fig. 5: Avalanche distribution for one-dimensional model of size $L=100$ with
$\alpha=0.7$. $S$, the size of avalanche, is the number of subsequent 
mutations below the threshold $0.728$. $P_{\rm aval}(S)$ denotes the 
distribution of $S$. The slope of the curve is about $0.774 \pm 0.001$.   

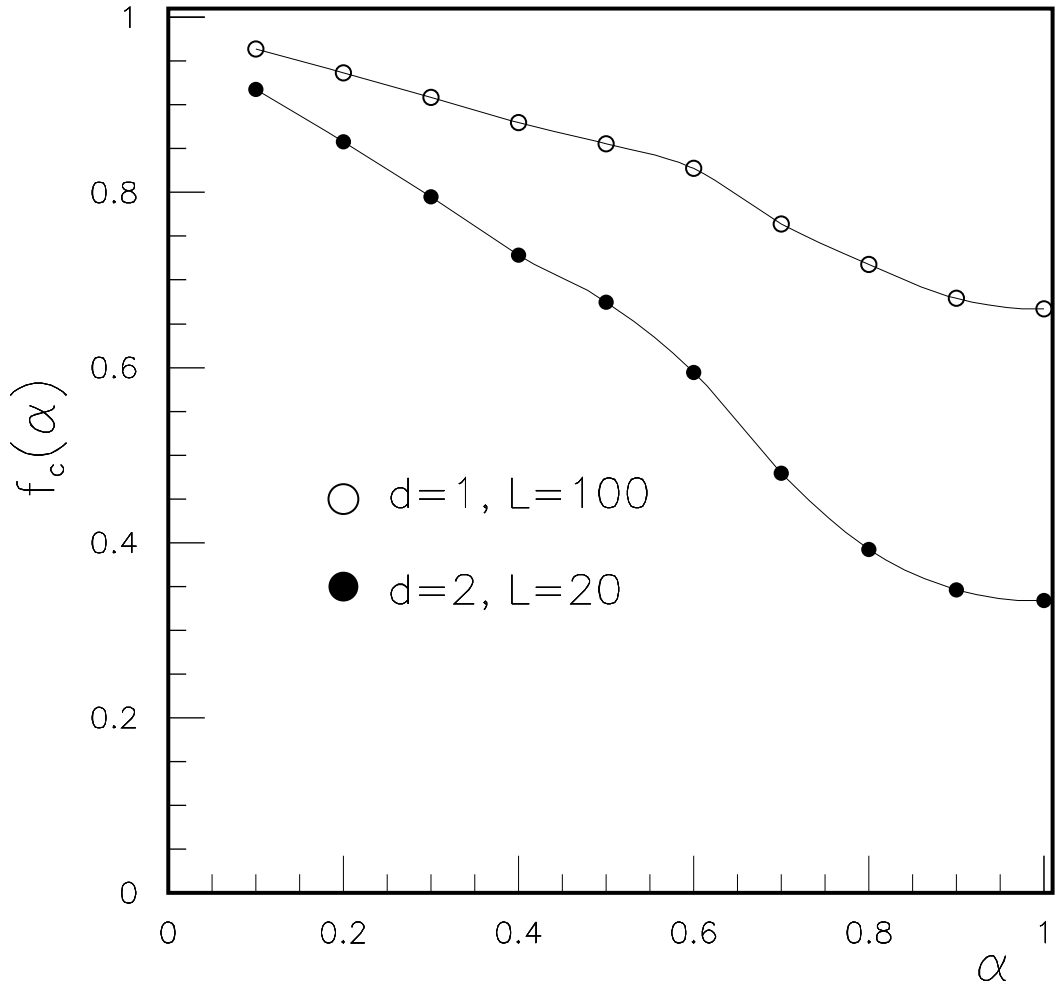
Fig. 6: Dependence of $f_{c}(\alpha)$ on $\alpha$ for one-dimensional
evolution models of size $L=100$ and two-dimensional models of size $L=20$.
 
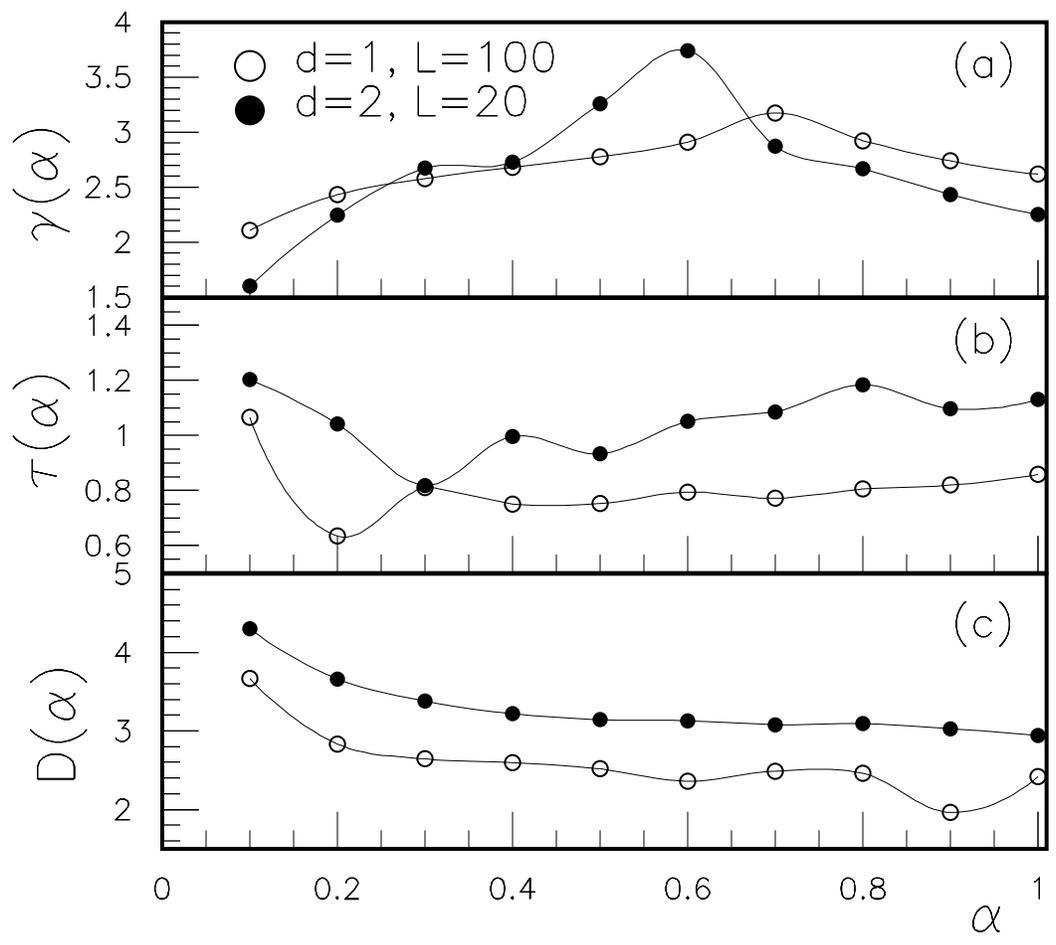
Fig. 7: (a) Dependence of $\gamma(\alpha)$ on $\alpha$ for one-dimensional
evolution models of size $L=100$ and two-dimensional models of size $L=20$.
(b) Dependence of $\tau(\alpha)$ on $\alpha$ for one-dimensional
evolution models of size $L=100$ and two-dimensional models of size $L=20$.
(c) Dependence of $D(\alpha)$ on $\alpha$ for one-dimensional
evolution models of size $L=100$ and two-dimensional models of size $L=20$.

\end{document}